\documentclass[conference]{IEEEtran}

\usepackage{url,hyperref,lineno,microtype,subcaption}
\usepackage{mathtools}
\usepackage{float}
\usepackage{graphicx}
\usepackage{subcaption}
\usepackage{booktabs,parskip}
\usepackage{algorithm}
\usepackage{algorithmic}
\usepackage{dsfont}
\usepackage[utf8]{inputenc}
\usepackage{dblfloatfix}
\usepackage{cite}
\usepackage{subcaption}
\usepackage{float}
\usepackage{graphics}
\usepackage{amsmath,amssymb,amsfonts}
\usepackage{algorithmic}
\usepackage{graphicx}
\usepackage{array}
\newcolumntype{P}[1]{>{\centering\arraybackslash}p{#1}}
\usepackage{textcomp}
\usepackage{xcolor}
\usepackage{ulem} 
\usepackage{textcomp}
\usepackage{mathtools}
\usepackage{authblk}
\def\BibTeX{{\rm B\kern-.05em{\sc i\kern-.025em b}\kern-.08em
    T\kern-.1667em\lower.7ex\hbox{E}\kern-.125emX}}

\usepackage[numbers]{natbib}
\usepackage{hyperref}
    
\begin{document}

\title{Model-based Estimation of AV-nodal Refractory Period and Conduction Delay Trends from ECG (Preprint) \\
{\footnotesize \textsuperscript{}}

}
\author[1,2]{Mattias Karlsson}
\author[3]{Pyotr G Platonov}
\author[4]{Sara R. Ulimoen}
\author[2]{Frida Sandberg}
\author[1]{Mikael Wallman}

\affil[1]{Department of Systems and Data Analysis, Fraunhofer-Chalmers Centre, Sweden}
\affil[2]{Department of Biomedical Engineering, Lund University, Sweden}
\affil[3]{Department of Cardiology, Clinical Sciences, Lund University, Sweden}
\affil[4]{Vestre Viken Hospital Trust, Department of Medical Research, B\ae rum Hospital, Norway}

\maketitle

\begin{abstract}
Atrial fibrillation (AF) is the most common arrhythmia, associated with significant burdens to patients and the healthcare system. The atrioventricular (AV) node plays a vital role in regulating heart rate during AF, but is often insufficient in regards to maintaining a healthy heart rate. Thus, the AV node properties are modified using rate-control drugs. Hence, quantifying individual differences in diurnal and short-term variability of AV-nodal function could aid in personalized treatment selection. \newline
\indent
~~~
This study presents a novel methodology for estimating the refractory period (RP) and conduction delay (CD) trends and their uncertainty in the two pathways of the AV node during 24 hours using non-invasive data. This was achieved using a network model together with a problem-specific genetic algorithm and an approximate Bayesian computation algorithm. Diurnal and short-term variability in the estimated RP and CD was quantified by the difference between the daytime and nighttime estimates and by the Kolmogorov-Smirnov distance between adjacent 10-minute segments in the 24-hour trends. \newline
\indent
~~~
Holter ECGs from 51 patients with permanent AF during baseline were analyzed, and the predictive power of variations in RP and CD on the resulting heart rate reduction after treatment with four rate control drugs was investigated. Diurnal variability yielded no correlation to treatment outcome, and no prediction of drug outcome was possible using the machine learning tools. However, a correlation between the short-term variability for the RP and CD in the fast pathway and resulting heart rate reduction during treatment with metoprolol ($\rho=0.48, p<0.005$ in RP, $\rho=0.35, p<0.05$ in CD) were found. \newline
\indent
~~~
The proposed methodology enables non-invasive estimation of the AV node properties during 24 hours, which may have the potential to assist in treatment selection. 

\end{abstract}


\section{Introduction} \label{Sec:intro}
\begin{figure*}[b]
    \centering
        \includegraphics[width=18cm]{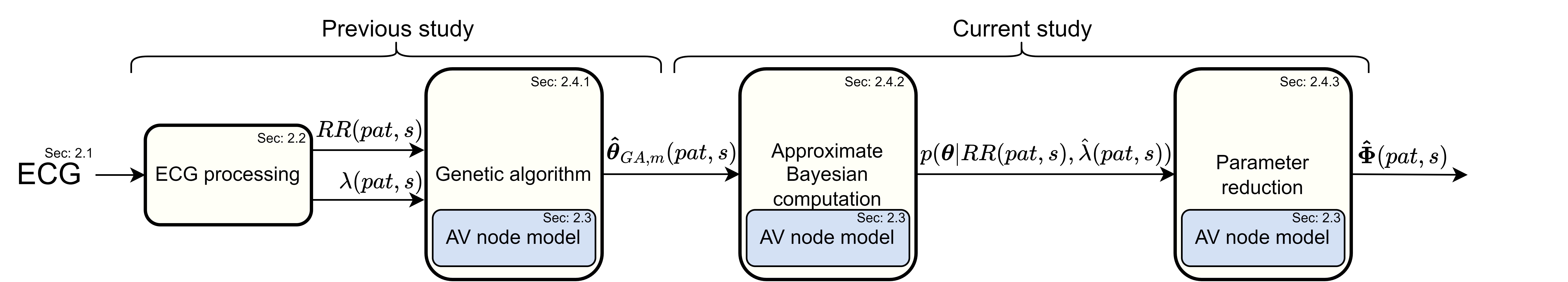}
        \caption{A schematic overview of the methodology, from ECG to estimations of the RP and CD. Previous study refers to \cite{karlsson2022ecg}. }
    \label{Figure1}
\end{figure*}

\noindent
Atrial fibrillation (AF) is the most common sustained cardiac arrhythmia and a significant burden for patients and the healthcare system \cite{ESC_Guidelines}. The prevalence of AF is currently estimated to be between 2 and 4\% worldwide \cite{benjamin2019heart}. In addition, the number of AF cases in the European Union is estimated to increase by 89\% between 2016 and 2060 \cite{di2019prevalence}. Atrial fibrillation is characterized by disorganized electrical activity in the atria, leading to rapid and irregular contraction, and is associated with an increased risk of mortality, predominantly due to heart failure or stroke \cite{andrew2013prevalence}.\newline
\indent ~~~
The atrioventricular (AV) node acts as the only electrical connection between the atria and ventricles and partly protects the ventricles from the rapid and irregular electrical activity in the atria during AF. It can be functionally divided into two pathways, the fast pathway (FP) and the slow pathway (SP), interconnected at the Bundle of His \cite{kurian2010anatomy}. The AV node either blocks an incoming impulse, based on its refractory period (RP), or sends it through with a delay, based on its conduction delay (CD). The AV node is thus the most essential part in regulating the heart rate during AF, and the RP and CD are the two most important properties of the AV node, deciding its filtering capability. \newline
\indent ~~~
The AV node during permanent AF is in many cases insufficient in regards to maintaining a healthy heart rate. Therefore, the AV node properties are often modified by treatment with rate control drugs, with $\beta$-blockers and calcium channel blockers recommended as first-line treatment \cite{ESC_Guidelines}. Common $\beta$-blockers for AF treatment are metoprolol and carvedilol, which block the $\beta 1$ receptors in the heart in order to reduce the effect of the sympathetic nervous system on the heart \cite{dorian2005antiarrhythmic}. Common calcium channel blockers are verapamil and diltiazem, which prevent the L-type calcium channels in the cardiac myocytes from opening in order to reduce conduction in the AV node \cite{eisenberg2004calcium}. However, due to the significant and poorly understood individual variability, the choice of drug is currently made empirically for each patient \cite{ESC_Guidelines}. This could lead to a prolonged time until successful treatment, and possibly result in a suboptimal final choice of drug. Since the two recommended first-line treatments have different physiological effects on the AV node, assessing the patient-specific properties of the AV node has the potential to assist in treatment selection. Specifically, we hypothesize that $\beta$-blockers would exhibit an increased effect (more reduced heart rate) when variations in the AV node properties are prominent since $\beta$-blockers reduce the effect of the sympathetic nervous system. \newline
\indent ~~~
The AV-node has previously been studied using several mathematical models based on invasive data from humans and animals \cite{billette1994dynamic,jorgensen2002mathematical,mangin2005effects,inada2009one,climent2011functional,mase2012nodal,mase2015dynamics,ryzhii2023compact}. However, in order for a model to be clinically applicable on an individual level, the model parameters should ideally be identifiable from non-invasive data, such as the ECG. A statistical model of the AV node with dual pathway physiology using the RR interval series and the atrial fibrillatory rate (AFR) for model fitting has been proposed \cite{corino2011atrioventricular,corino2013atrioventricular,henriksson2015statistical}. However, the model lumps RP and CD together, limiting its interpretability. \newline
\indent ~~~
We have previously proposed a network model of the AV node \cite{karlsson2021non} together with a framework for continuously estimating twelve model parameters describing the RP and CD in the two pathways from 24-hour Holter ECG \cite{karlsson2022ecg}. Although promising, the characterization of the AV node was still limited by the number of model parameters and their intrinsic complex dependencies, where a large change in the model parameters could result in a very small change in the RP or CD, thus, making their interpretation a non-trivial task. For a modeling approach to gain acceptance in a clinical context, the outcome should be readily interpretable by medical professionals; a fact that has become especially relevant with the increasing use of advanced modeling and machine learning techniques \cite{teng2022survey,trayanova2021machine}. Additionally, in \cite{karlsson2022ecg}, a version of Sobol's method was applied to quantify uncertainty in the parameter estimates. However, these uncertainty estimates were not directly interpretable as probabilities and could thus only be used as a relative measure between the model parameters, between patients, or between different times of the day. When the extent of the uncertainty is unknown, uncertain estimates have the potential to mislead decision-making processes or further analysis of the trends. A proper quantification of the uncertainty is thus advantageous in order to fully understand the estimates.\newline
\indent ~~~
In the present study, we propose a novel methodology for estimating the RP and CD of both pathways of the AV node and the associated uncertainty continuously over 24 hours. The methodology comprises a genetic algorithm (GA) for initial model parameter estimation and an approximate Bayesian computation (ABC) algorithm to refine the estimates, together with a simulation approach to map model parameters to RP and CD in order to increase interpretability. In addition to refining the estimates, the ABC algorithm provides samples from the Bayesian posterior distribution of the AV node properties, hereafter denoted the posterior, enabling proper quantification of the uncertainty of the estimated properties. We employ these novel tools in an exploratory manner to analyze Holter ECGs from 51 patients during baseline in combination with their respective drug responses to identify potential markers for differences in drug response. Specifically, we analyze the correlation between diurnal and short-term variability and drug outcomes, as well as train several machine learning models to predict drug outcomes.

\section{Materials and Methods} \label{Sec:method}

\begin{figure*}[b]
    \centering
        \includegraphics[width=18cm]{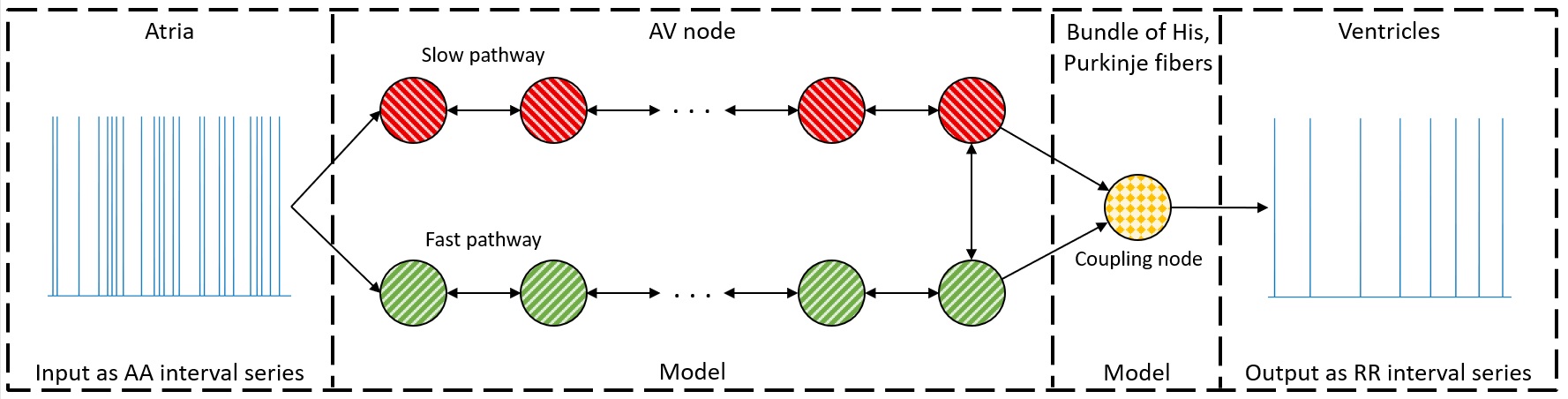}
    \caption{A schematic representation of the network model where the yellow node represents the coupling node, the red nodes the SP, the green nodes the FP, and arrows the direction for impulse conduction. For readability, only a subset of the 21 nodes is shown \cite{karlsson2021non}.}
    \label{Figure2}
\end{figure*}

The overall method for assessing the RP and CD of the two pathways in the AV node for each patient ($pat$) can be divided into four stages, as shown in Figure \ref{Figure1}. Firstly, 24-hour Holter ECGs are processed to extract RR interval series and AFR trends, divided into ten-minute segments ($s$) with a 50\% overlap, as described in Sections \ref{Sec:3_1} and \ref{Sec:3_2}. Secondly, the parameters for the network model of the AV node, described in Section \ref{Sec:3_3}, are fitted to the RR interval series and AFR in each segment using a problem-specific dynamic GA as described in Section \ref{Sec:3_4_GA}. The GA-derived estimates are subsequently used as inputs to an ABC algorithm to refine the estimates and estimate the posterior of the model parameters, as described in Section \ref{Sec:3_4_ABC}. These model parameter estimates are finally used to simulate data with the model while tracking the RP and CD used for the two pathways, as described in Section \ref{Sec:3_4_reduction}. This results in a distribution of the RP and CD in the FP and the SP for each ten-minute segment. Finally, the possibility to predict treatment outcomes using the estimated distributions is evaluated, as described in Section \ref{sec:pred}.

\subsection{ECG Data}\label{Sec:3_1}
\noindent Data from the Rate Control in Atrial Fibrillation (RATAF) study, a randomized, investigator-blind, crossover study, approved by the regional ethics committee and the Norwegian Medicines Agency and conducted in accordance with the Helsinki Declaration, is analyzed in this study \cite{ulimoen2013comparison}. Specifically, 24-hour ambulatory ECGs from 60 patients (mean age 71 $\pm$ 9 years, 18 women) with permanent AF, no heart failure, or symptomatic ischemic heart disease, recorded during baseline, are used for the estimation of patient-specific AV node properties. In addition to the baseline ECG, the relative change in the 24-hour average heart rate ($\Delta HR$) for treatment with the two calcium channel blockers verapamil and diltiazem and the two $\beta$-blockers metoprolol and carvedilol are used to evaluate the therapeutic implications of the estimated AV node properties. The calculation of $\Delta HR$ is based on the RR interval series extracted from the ECG, as explained in Section \ref{Sec:3_2}. 

\subsection{ECG Processing} \label{Sec:3_2}
\noindent The RR interval series is extracted from the ECG for each patient and divided into ten-minute segments with a 50\% overlap ($\boldsymbol{RR}(pat,s)$), where RR intervals following and preceding QRS-complexes with deviating morphology are excluded from the series \cite{lagerholm2000clustering}. Segments with excessive noise can lead to a large number of undetected beats and thus an unrealistically low heart rate. Hence, each ten-minute segment is divided into minute-long non-overlapping intervals, and the whole ten-minute segment is excluded from further analysis if any one-minute interval has fewer than 20 detected beats. Patients with RR interval series with a total duration shorter than 12 h are excluded from further analysis. The RR interval series corresponding to the four rate control drugs are calculated equivalently. \newline
\indent ~~~
Spatiotemporal QRST cancellation is employed to extract the f-waves from the ECG \cite{stridh2001spatiotemporal}. Subsequently, the fundamental frequency of the extracted f-waves is tracked using a hidden Markov model-based method to extract an AFR trend for each patient with a resolution of one minute \cite{sandberg2008frequency}. For time segments where the AFR could not be obtained due to excessive noise, but the RR interval series could, the AFR is set to the closest observed AFR value.

\begin{table*}[b]
\caption{Parameter ranges for the GA and the ABC PMC algorithm. \label{Table1}
}
\centering
\label{Tab:Ranges}
\begin{tabular}{|c|c|c|c|c|c|}
\hline
Parameters &  $R_{min}^{FP}, R_{min}^{SP}$ & $\Delta R^{FP}, \Delta R^{SP}$ & $D_{min}^{FP}, D_{min}^{SP}$ & $\Delta D^{FP}, \Delta D^{SP}$ & $\tau_R^{FP},\tau_R^{SP},\tau_D^{FP},\tau_D^{SP}$ \\
\hline
 GA (ms) & [100, 1000] & [0, 1000] & [2, 50] & [0, 100] & [25, 500]  \\
 ABC (ms) & [30, 1300] & [0, 1300] & [0.1, 80] & [0, 130] & [10, 700]  \\
\hline
\end{tabular} 
\end{table*}

\subsection{Network Model of the AV Node} \label{Sec:3_3}
\noindent Our network model of the AV node, introduced in \cite{karlsson2021non}, describes the AV node as two pathways (the SP and the FP) comprising 10 nodes each. These two pathways are connected by a coupling node, as illustrated in Figure \ref{Figure2}. Each pathway node corresponds physiologically to a localized section of the respective pathway, and the coupling node corresponds to the Purkinje fibers and Bundle of His.\newline
\indent ~~~
Atrial impulses are modeled by a Poisson process with mean arrival rate $\lambda$. The impulses are assumed to reach the first nodes of SP and FP simultaneously. Each network node can be either in a refractory state or in a non-refractory state. A node in its refractory state will block incoming impulses, and a node in its non-refractory state will transmit an incoming impulse to all adjacent nodes with an added conduction delay before entering its refractory state. The RP ($R_i(n)$) and CD ($D_i(n)$) for node $i$ are updated for each incoming impulse $n$ according to Equations \ref{eq:R}, \ref{eq:D}, and \ref{eq:TDI},

\begin{equation} \label{eq:R}
    R_i(n) = R_{min} + \Delta R(1-e^{-\tilde{t}_{i}(n)/\tau_R})
\end{equation}
\begin{equation} \label{eq:D}
    D_i(n) = D_{min} + \Delta D e^{-\tilde{t}_{i}(n)/\tau_D},
\end{equation}
\begin{equation} \label{eq:TDI}
    \tilde{t}_{i}(n) = t_i(n) - ( t_i(n-1) + R_i(n-1) ),
\end{equation}

where, $\tilde{t}_i(n)$ is the diastolic interval preceding impulse $n$ and $t_i(n)$ is the arrival time of impulse $n$ at node $i$. When $\tilde{t}_{i}(n) < 0$, the node is in its refractory state and will block incoming impulses. All parameters are fixed for each pathway, resulting in three model parameters for the RP in the FP ($R_{min}^{FP},\ \Delta R^{FP},\ \tau_R^{FP}$); three model parameters for the CD in the FP ($D_{min}^{FP},\ \Delta D^{FP},\ \tau_D^{FP}$); three model parameters for the RP in the SP ($R_{min}^{SP},\ \Delta R^{SP},\ \tau_R^{SP}$); three model parameters for the CD in the SP ($D_{min}^{SP},\ \Delta D^{SP},\ \tau_D^{SP}$). These twelve model parameters constitute the mode parameter vector $\boldsymbol{\theta}$. In addition, the RP in the coupling node is fixed to the mean of the ten shortest RR intervals in the data, and its CD is fixed at 60 ms \cite{karlsson2021non}. 

\label{Sec:Model}

\subsection{Parameter Estimation}
For each ten-minute segment, the mean arrival rate for the Poisson process $\lambda$ is estimated as the mean of the AFR trend ($\hat{\lambda}(pat,s)$), and the model parameters $\boldsymbol{\hat{\theta}}(pat,s)$ are estimated using a GA together with an ABC algorithm. \newline
\indent ~~~
An error function ($\epsilon$) based on the Poincaré plot, i.e., a scatter plot of successive pairs of RR intervals, is used to quantify the difference between $\boldsymbol{RR}(pat,s)$ and a simulated RR interval series ($\boldsymbol{\Tilde{RR}}$). The successive pairs of RR intervals for $\boldsymbol{RR}(pat,s)$ and $\boldsymbol{\Tilde{RR}}$ are placed in two-dimensional bins covering the interval between 250 and 1800 ms in steps of 50 ms, resulting in $K$ = 961 bins, which we refer to as the Poincaré histogram. The error function, based on the work presented in \cite{karlsson2021non}, is computed according to Equation \ref{Eq:poincare},

\begin{equation} \label{Eq:poincare}
    \epsilon = \frac{1}{K} \sum_{k=1}^{K} \frac{\Big(x_{k} - \frac{1}{t_{norm}} \Tilde{x}_k\Big)^2}{\sqrt{x_k}},
\end{equation}

where $x_k$ and $\Tilde{x}_k$ are the numbers of RR intervals in the $k$-th bin of $\boldsymbol{RR}(pat,s)$ and $\boldsymbol{\Tilde{RR}}$, respectively. Additionally, $t_{norm}$ acts as a normalizing constant and is calculated as the duration of $\boldsymbol{\Tilde{RR}}$ divided by the duration of $\boldsymbol{RR}(pat,s)$.

\subsubsection{Genetic Algorithm} \label{Sec:3_4_GA}
A problem-specific dynamic GA based on the work presented in \cite{karlsson2022ecg} is used to get an initial estimate of $\boldsymbol{\theta}(pat,s)$ in every segment. This results in an estimate denoted as $\boldsymbol{\hat{\theta}}^{GA}_{m}(pat,s)$, where $m$ denotes the $m$-th fittest individual in the population after completion of the GA, i.e. the individual with the $m$-th lowest $\epsilon$. The hyper-parameters in the algorithm are tuned during the optimization using the difference between the Poincaré histograms in pairs of consecutive segments ($\Delta P$) \cite{karlsson2022ecg}. This difference is calculated using Equation \ref{Eq:poincare} with $x_k$ and $\Tilde{x}_k$ as the number of RR intervals in each bin of the current segment and the following one, respectively. \newline
\indent ~~~
The GA uses a population of 300 individuals, where each individual is a model parameter vector $\boldsymbol{\theta}$. The algorithm uses tournament selection, a two-point crossover, and creep mutation. To avoid premature convergence and to increase performance, immigration through replacement of the least-fit individuals in the population is performed, following the work in \cite{karlsson2022ecg}. Furthermore, $\Delta P$ is used to determine the number of generations that the GA runs before moving to the next data segment, between two and seven. The initialization of individuals is done using latin hypercube sampling within the ranges given in Table \ref{Table1}. These values also act as boundaries for the model parameters in the GA. For further details about the algorithm, see \cite{karlsson2022ecg}.

\subsubsection{Approximate Bayesian Computation} \label{Sec:3_4_ABC}
To estimate the posterior $p(\boldsymbol{\theta}|\boldsymbol{RR}(pat,s),\hat{\lambda}(pat,s))$, an approximate Bayesian computation population Monte Carlo sampling (ABC PMC) algorithm is used \cite{turner2012tutorial}. The pseudo-code for the problem-specific ABC PMC is shown in Algorithm \ref{Alg}. The ABC PMC uses a set of $N_{p} = 100$ particles to estimate the posterior in each RR segment independently, which are updated iteratively for eight iterations ($j$). Each particle corresponds to a model parameter vector, denoted $\boldsymbol{\hat{\theta}}^{ABC}_{v, j}$, where $v$ corresponds to the $v$-th particle for the $j$-th iteration. The algorithm is sped up by utilizing the results from the GA to create the initial population. To construct the initial population, twenty particles are drawn from five different normal distributions, $\mathcal{N}(\boldsymbol{\hat{\theta}}^{GA}_{1}, \boldsymbol{\Sigma}_{init})$, $\mathcal{N}(\boldsymbol{\hat{\theta}}^{GA}_{2}, \boldsymbol{\Sigma}_{init})$, $\mathcal{N}(\boldsymbol{\hat{\theta}}^{GA}_{3}, \boldsymbol{\Sigma}_{init})$, $\mathcal{N}(\boldsymbol{\hat{\theta}}^{GA}_{4}, \boldsymbol{\Sigma}_{init})$, and $\mathcal{N}(\boldsymbol{\hat{\theta}}^{GA}_{5}, \boldsymbol{\Sigma}_{init})$, where the covariance matrix $\boldsymbol{\Sigma}_{init} = $ Cov$(\boldsymbol{\hat{\theta}}^{GA}_{1:25})$ and $1:25$ denotes $[1,2,...,25]$ for convenience. During each iteration, each particle has a probability of being chosen based on an assigned weight, computed according to Equation \ref{Eq:weights} \cite{beaumont2009adaptive}

\begin{equation} \label{Eq:weights}
    \boldsymbol{w}_{v, j} = \big( \sum^{N_p}_{k = 1} \boldsymbol{w}_{k, j-1} \mathcal{N}(\boldsymbol{\hat{\theta}}^{ABC}_{k, j-1} | \boldsymbol{\hat{\theta}}^{ABC}_{v,j}, \boldsymbol{\Sigma}_{j-1})\big)^{-1},
\end{equation}

where $\boldsymbol{w}_{v, j}$ is the weight for the $v$-th particle in the $j$-th iteration and $\mathcal{N}(\boldsymbol{\hat{\theta}}^{ABC}_{k, j-1} | \boldsymbol{\hat{\theta}}^{ABC}_{v,j}, \boldsymbol{\Sigma}_{j-1})$ is the probability of $\boldsymbol{\hat{\theta}}^{ABC}_{k, j-1}$ given the normal distribution with mean $\boldsymbol{\hat{\theta}}^{ABC}_{v,j}$ and covariance $\boldsymbol{\Sigma}_{j-1}$, where $\boldsymbol{\Sigma}_{j} = 2$Cov$(\boldsymbol{\hat{\theta}}^{ABC}_{1:N_p, j})$. Furthermore, the chosen particle ($\boldsymbol{\theta}^*$) is perturbed to create a proposal particle ($\boldsymbol{\theta}^{**}$) using a transition kernel set as $\mathcal{N}$$(0,\boldsymbol{\Sigma}_{j})$ \cite{beaumont2009adaptive}. The model is used to simulate data using $\boldsymbol{\theta}^{**}$ to calculate an associated proposal error ($\epsilon^{**}$) according to Equation \ref{Eq:poincare}. If $\epsilon^{**}$ is lower than a set threshold ($T_{j}$), $\boldsymbol{\theta}^{**}$ is accepted and used in the next iteration of the algorithm; if not, a new particle is chosen and perpetuated to create a new proposal particle. Note that the boundaries for the ABC PMC algorithm are more inclusive compared to the GA to accommodate the full width of the estimated posteriors, as shown in Table \ref{Table1}. A proposal particle outside the boundaries is always rejected. The next iteration starts when $N_{p}$ new proposal particles have been accepted, and $\boldsymbol{w}_{v, j}$, $T_{j}$, and $\boldsymbol{\Sigma}_{j}$ are then updated. The threshold changes based on the results from the GA; where $T_{1} = \boldsymbol{\hat{\theta}}^{GA}_{10}(pat,s)$, $T_{2} = \boldsymbol{\hat{\theta}}^{GA}_{8}(pat,s)$, $T_{3} = \boldsymbol{\hat{\theta}}^{GA}_{5}(pat,s)$, $T_{4} = \boldsymbol{\hat{\theta}}^{GA}_{3}(pat,s)$, and $T_{5:8} = \boldsymbol{\hat{\theta}}^{GA}_{1}(pat,s)$. Hence, after the eighth iteration, the $\epsilon$ for all particles is lower than the $\epsilon$ for the fittest individual found by the GA. Thus, the final population is assumed to be $N_{p}$ samples from $p(\boldsymbol{\theta}|\boldsymbol{RR}(pat,s),\hat{\lambda}(pat,s))$.


\begin{algorithm*}[b]
\caption{Calculate $ p(\boldsymbol{\theta}|\boldsymbol{RR},\hat{\lambda} )$, given $\boldsymbol{RR}$, $\hat{\lambda}$, the model $\boldsymbol{\Tilde{RR}} \sim$ Model($\boldsymbol{\theta}$, $\hat{\lambda}$), the threshold $T_{j}$, and the initial estimates\vspace{1.5pt} \newline $\boldsymbol{\hat{\theta}}^{GA}$. The indication $(pat,s)$ is omitted to avoid redundancy.}
\begin{algorithmic} \label{Alg}
\STATE At iteration $j = 1$, set the initial population
\STATE Set a counter $c = 1$
\FOR{$1 \leq u \leq 5$}
\FOR{$1 \leq q \leq \frac{N_p}{5}$}
\STATE Set $\boldsymbol{\hat{\theta}}^{ABC}_{c, 1} \leftarrow \mathcal{N}(\boldsymbol{\hat{\theta}}^{GA}_{u}, \boldsymbol{\Sigma}_{init}) $
\STATE Set initial weights  $\boldsymbol{w}_{c, 1} \leftarrow \frac{1}{N_p}$
\STATE Update counter $c = c + 1$
\ENDFOR
\ENDFOR
\STATE Set the initial covariance for the transition kernel $\boldsymbol{\Sigma}_1 \leftarrow$ 2Cov$(\boldsymbol{\hat{\theta}}^{ABC}_{1:N_p, 1})$
\STATE At iteration $j>1$
\FOR{$2 \leq j \leq 8$}
\FOR{$1 \leq v \leq N_p$}
\STATE Set $\epsilon^{**}$ = inf 
\WHILE{$\epsilon^{**} > T_j$}
\STATE Sample one proposal particle from previous iteration $\boldsymbol{\theta}^* \sim \boldsymbol{\hat{\theta}}^{ABC}_{1:N_p,j-1}$ with probability $\boldsymbol{w}_{1:N_p, j-1}$
\STATE Perturb $\boldsymbol{\theta}^*$ by sampling $\boldsymbol{\theta}^{**} \sim \mathcal{N}(\boldsymbol{\theta}^*, \boldsymbol{\Sigma}_{j-1})$
\STATE Simulate data $\boldsymbol{\Tilde{RR}}$ from $\boldsymbol{\theta}^{**}$: $\boldsymbol{\Tilde{RR}} \sim$ Model($\boldsymbol{\theta}^{**}, \hat{\lambda}$)
\STATE Calculate $\epsilon^{**}$ from Equation \ref{Eq:poincare} using $\boldsymbol{\Tilde{RR}}$ and $\boldsymbol{RR}$
\ENDWHILE
\vspace{2pt}
\STATE Set $\boldsymbol{\hat{\theta}}^{ABC}_{v,j} \leftarrow \boldsymbol{\theta}^{**}$ 
\STATE Update the weight $\boldsymbol{w}_{v, j} \leftarrow \big( \sum^{N_p}_{k = 1} \boldsymbol{w}_{k, j-1} P(\boldsymbol{\hat{\theta}}^{ABC}_{k, j-1}|N~(\boldsymbol{\hat{\theta}}^{ABC}_{v,j}, \boldsymbol{\Sigma}_{j-1})) \big)^{-1} $ (Equation \ref{Eq:weights})
\ENDFOR
\STATE Update the covariance for the transition kernel $\boldsymbol{\Sigma}_{j} \leftarrow 2$Cov$(\boldsymbol{\hat{\theta}}^{ABC}_{1:N_p, j})$
\ENDFOR
\end{algorithmic}
\end{algorithm*}

\subsubsection{Parameter Reduction} \label{Sec:3_4_reduction}
The posterior estimate of the parameter vector $\boldsymbol{\theta}(pat,s)$ is obtained using the resulting $N_p$ samples ($\boldsymbol{\hat{\theta}}^{ABC}_{1:N_p, 8}(pat,s)$) from the ABC PMC algorithm. Each $\boldsymbol{\hat{\theta}}^{ABC}_{1:N_p, 8}(pat,s)$ is utilized within the model together with the associated $\hat{\lambda}(pat,s)$ to simulate a ten-minute RR interval series. For each simulation, $R_i(n)$ and $D_i(n)$ are stored for each activation $n$ in each pathway node $i$ and used as the sample distribution of the RP and CD for the SP and the FP, respectively. The samples from these four distributions, denoted as $\boldsymbol{\hat{\Phi}}(pat,s) = [\boldsymbol{R}^{FP}(pat,s), \boldsymbol{R}^{SP}(pat,s), \boldsymbol{D}^{FP}(pat,s), \boldsymbol{D}^{SP}(pat,s)]$, serves as a translation from the twelve model parameters $\boldsymbol{\hat{\theta}}$ to four more interpretable AV node properties $\boldsymbol{\hat{\Phi}}$, taking into account not only the model parameters but also the mean AFR associated with the current RR-segment.\newline
\indent ~~~
To quantify these distributions, their corresponding empirical probability density functions are computed using the MATLAB function ksdensity (MATLAB R2022b) with default bandwidth. From the empirical probability density functions, the maxima are obtained, denoted $\boldsymbol{\hat{\phi}}_{max}(pat,s) = [R^{FP}_{max}(pat,s), R^{SP}_{max}(pat,s), D^{FP}_{max}(pat,s), D^{SP}_{max}(pat,s)]$. In addition, the 5th percentile and the 95th percentile are obtained from $\boldsymbol{\hat{\Phi}}(pat,s)$, denoted $\boldsymbol{\hat{\phi}}_{5}(pat,s) = [R^{FP}_{5}(pat,s), R^{SP}_{5}(pat,s), D^{FP}_{5} (pat,s), D^{SP}_{5}(pat,s)]$, and $\boldsymbol{\hat{\phi}}_{95}(pat,s) = [R^{FP}_{95}(pat,s), R^{SP}_{95}(pat,s), D^{FP}_{95}(pat,s), \\ D^{SP}_{95}(pat,s)]$, respectively. Furthermore, the number of impulses traveling through the FP and SP ($N_{FP}$ and $N_{SP}$, respectively) is stored, and the ratio is denoted as 
\begin{equation}
\centering
    SP_{ratio}(pat,s) = \frac{N_{SP}(pat,s)}{N_{FP}(pat,s)+N_{SP}(pat,s)}.
\end{equation}
\newline
\indent ~~~
The patient-specific diurnal variability ($\Delta DV$) in the AV node properties is quantified by the average value of $\boldsymbol{\hat{\phi}}_{max}$ during daytime (9:00 A.M. to 9:00 P.M) divided by the average value of $\boldsymbol{\hat{\phi}}_{max}$ during nighttime (2:00 A.M. to 6:00 A.M). In addition, the patient-specific short-term variability in the AV node properties is quantified by the average Kolmogorov-Smirnov distance ($\overline{\Delta KS}$) between consecutive segments of $\boldsymbol{\hat{\Phi}}$ during the full 24-hour (8:00 A.M to 8:00 A.M). The Kolmogorov-Smirnov distance represents the maximal separation between the empirical cumulative distribution functions between consecutive segments \cite{massey1951kolmogorov}. 

\label{Sec:Fit}

\begin{figure*}[b]
    \centering
        \includegraphics[width=18cm]{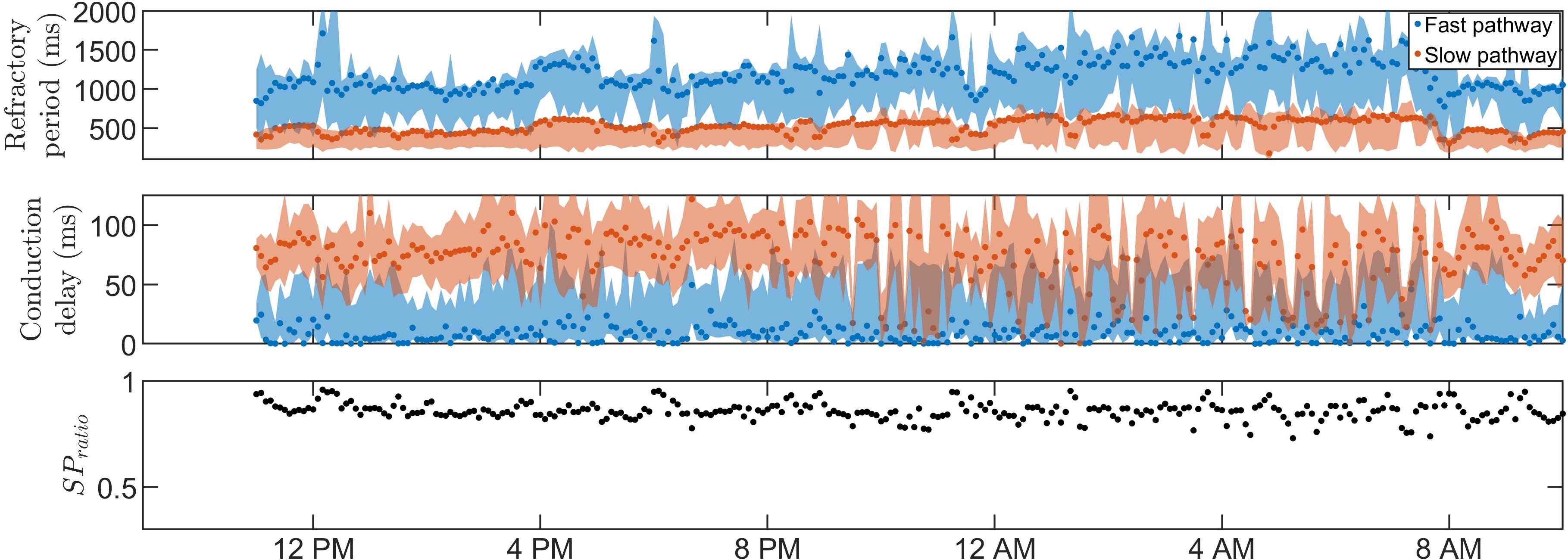}
    \caption{The estimated RF (top) and CD (middle) for $\boldsymbol{\hat{\phi}}_{max}(pat,s)$ (dotted) as well as $\boldsymbol{\hat{\phi}}_{5}(pat,s)$ and $\boldsymbol{\hat{\phi}}_{95}(pat,s)$ (filled) for the FP (blue) and SP (red), as well as the SP ratio (bottom) are shown for patient A, marked with a black circle in Figure \ref{Figure6}. }
    \label{Figure3}
\end{figure*}

\subsection{Prediction of Treatment Outcome} \label{sec:pred}
The predictive power of the estimates $\boldsymbol{\hat{\Phi}}$, $\boldsymbol{\hat{\phi}}_{5}$, $\boldsymbol{\hat{\phi}}_{95}$, $\boldsymbol{\hat{\phi}}_{max}$, and ${SP}_{ratio}$ in relation to $\Delta HR$ for the different rate control drugs is evaluated in three ways; by analyzing the correlation between the diurnal and short-time variability and $\Delta HR$; by training a feature-based regression model on statistical properties of the trends to predict $\Delta HR$; and by training a convolutional neural network on the trends to predict $\Delta HR$. \newline
\indent ~~~
To quantify the correlation between diurnal and short-term variability in the AV node properties and $\Delta HR$ after treatment with the four rate control drugs, Spearman's rank correlation is used. Due to the exploratory nature of the study, no hypothesis test is performed and hence no correction of p-values is applied \cite{perneger1998s,althouse2016adjust}. \newline
\indent ~~~
Three different feature-based regression models (linear regression, random forest \cite{breiman2001random}, and k-nearest neighbor \cite{cover1967nearest}) are trained on 66 statistical properties of the trends. These statistical properties are; the mean $\pm$ std of the four AV node properties $\boldsymbol{\hat{\phi}}_{max}$ during daytime (8 properties), during nighttime (8 properties), and the full 24-hour (8 properties); the mean $\pm$ std of the 90\% credibility region -- calculated as the difference between $\boldsymbol{\hat{\phi}}_{5}$ and $\boldsymbol{\hat{\phi}}_{95}$ -- during daytime (8 properties), nighttime (8 properties), and the full 24-hour (8 properties); the mean $\pm$ std of the ${SP}_{ratio}$ during daytime (2 properties), nighttime (2 properties), and the full 24-hour (2 properties); $\Delta DV$ in the four AV node properties (4 properties); the short-term variability in the four AV node properties (4 properties); as well as the age, gender, weight, and height of the patient. \newline
\indent ~~~
Deep learning approaches have achieved the current state-of-the-art performance for time-series classification and regression \cite{ismail2019deep}. Hence, the prediction of $\Delta HR$ for the different rate control drug is evaluated using the time series for $\boldsymbol{\hat{\phi}}_{5}$, $\boldsymbol{\hat{\phi}}_{95}$, $\boldsymbol{\hat{\phi}}_{max}$, ${SP}_{ratio}$, AFR, and the RR interval series as an input to three convolutional neural networks with different architectures, based on only fully connected layers \cite{wang2017time}, the ResNet architecture \cite{wang2017time}, and the Inception architecture \cite{ismail2020inceptiontime}, respectively. To incorporate the age, gender, weight, and height of the patients, the last fully connected layer of the networks is modified to also include these properties as input neurons. The networks were trained using the tsai library \cite{tsai}, with the Adam solver \cite{kingma2014adam} and the Huber loss \cite{huber1992robust}. Leave-one-out cross-validation is used, so that the network is trained on data from all but one patient and tested on the left-out patient. The average mean square error (MSE) of the predicted and true $\Delta HR$ for the whole population is calculated and compared between approaches.

\label{Sec:Pred}

\begin{figure*}[b]
    \centering
        \includegraphics[width=18cm]{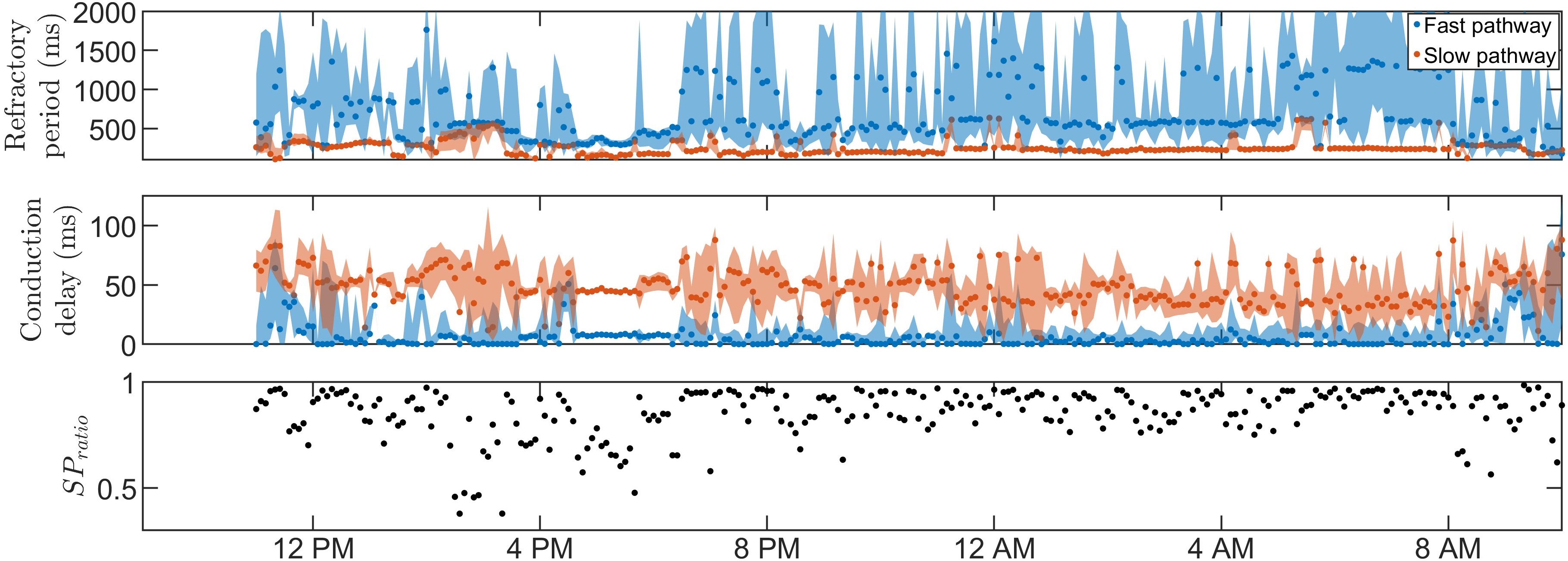}
    \caption{The estimated RF (top) and CD (middle) for $\boldsymbol{\hat{\phi}}_{max}(pat,s)$ (dotted) as well as $\boldsymbol{\hat{\phi}}_{5}(pat,s)$ and $\boldsymbol{\hat{\phi}}_{95}(pat,s)$ (filled) for the FP (blue) and SP (red), together with the SP ratio (bottom) are shown for patient B, marked with a red circle in Figure \ref{Figure6}. }
    \label{Figure4}
\end{figure*}

\vspace{-4pt}
\section{Results} \label{Sec:result}
\label{Sec:result}
\noindent As described in Section \ref{Sec:3_1}, this study is based on a population of 60 patients. However, due to excessive noise, some patients are excluded from analysis, as described in Section \ref{Sec:3_2}, resulting in a total of 51 patients. In addition, excessive noise in the ECG during treatment with the four rate control drugs leads to missing values for $\Delta HR$ for some patients. Thus, of the remaining 51 patients at baseline, two lack data for verapamil, three lack data for diltiazem, two lack data for metoprolol, none lack data for carvedilol, and one lacks data for both verapamil and metoprolol. The mean $\pm$ standard deviation of $\Delta HR$ in the population are $19\% \pm 23\%$ for verapamil; $24\% \pm 18\%$ for diltiazem, $17\% \pm 18\%$ for metoprolol; and $11\% \pm 6\%$ for carvedilol. \newline
\indent ~~~
\subsection{Parameter Trends} \label{Res:trends}
\begin{figure*}[b]
    \centering
        \includegraphics[width=18cm]{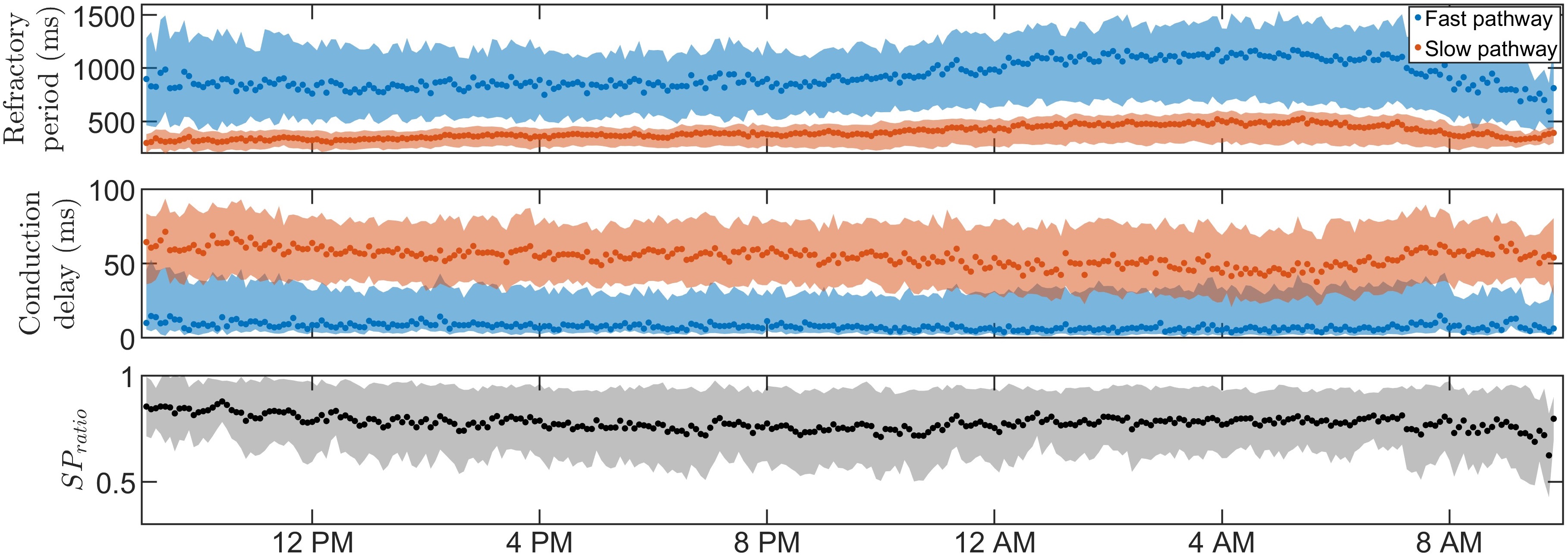}
    \caption{The average RF (top) and CD (middle) for $\boldsymbol{\hat{\phi}}_{max}(pat,s)$ (dotted) as well as $\boldsymbol{\hat{\phi}}_{5}(pat,s)$ and $\boldsymbol{\hat{\phi}}_{95}(pat,s)$ (filled) for the FP (blue) and SP (red), together with the mean (black, dotted) and standard deviation (black, filled) of the SP ratio (bottom). }
    \label{Figure5}
\end{figure*}

The 24-hour trends of $\boldsymbol{\hat{\phi}}_{max}(pat,s)$, $\boldsymbol{\hat{\phi}}_{5}(pat,s)$, and $\boldsymbol{\hat{\phi}}_{95}(pat,s)$ for two patients, denoted A and B, are presented in Figure \ref{Figure3} and \ref{Figure4}. Figure \ref{Figure3} shows a low short-term variability in the RP and CD in both pathways for patient A ($\overline{\Delta KS} = [0.27, 0.19, 0.24, 0.33]$ for $R^{FP}$, $R^{SP}$, $D^{FP}$, and, $D^{SP}$), whereas patient B in Figure \ref{Figure4} has a larger short-term variability ($\overline{\Delta KS} = [0.41, 0.55, 0.40, 0.40]$). Conduction mainly occurs through the SP in both patients, as indicated by an $SP_{ratio}$ over 0.5, which results in a wider credibility region in the $R^{FP}$ compared to the $R^{SP}$. However, for patient B, there are segments where the FP is more prevalent, e.g. between 5 PM and 6 PM. In these segments, the RP and CD have a very low variability indicating a stationary behavior of the AV node. A notable shift in RP occurs at 8 AM for patient A, probably as a response to waking up from sleep, resulting in a clear change in autonomic regulation. No notable difference between the average $R^{FP}$, $R^{SP}$, and $D^{FP}$ during daytime and during nighttime could be seen for patient A, with a slight difference in $D^{SP}$ ($\Delta DV$ $= [0.80, 0.81, 0.99, 1.39]$). For patient B, only $D^{FP}$ showed a notable difference ($\Delta DV$ $= [0.81, 0.92, 2.60, 1.19]$). \newline
\indent ~~~
Similar observations can be made for the whole population, as displayed in Table \ref{Table2}, which includes the mean and standard deviation of $\boldsymbol{\hat{\phi}}_{max}(pat,s)$, the 95\% credibility region, and $\overline{\Delta KS}$, during daytime, nighttime, and during 24 hours, as well as $\Delta DV$, for the RP and CD in the FP and the SP for all patients. For convenience, the total CD, calculated by multiplying the CD for one node by ten, is listed. From Table \ref{Table2}, it is evident that the RP on average is higher and the CD is lower during nighttime compared to daytime, probably linked to the lower heart rate during sleep and/or circadian autonomic variations. Figure \ref{Figure5} illustrates the population average trends of $\boldsymbol{\hat{\phi}}_{max}(pat,s)$, $\boldsymbol{\hat{\phi}}_{5}(pat,s)$, and $\boldsymbol{\hat{\phi}}_{95}(pat,s)$. To reduce the influence of outliers, only segments containing data from over 20\% of the population are shown. A distinct separation between RP and CD of the two pathways exists, indicating different functionality. Additionally, the credibility region for the $R^{FP}$ is larger compared to the $R^{SP}$. Moreover, the credibility region for $D^{FP}$, in proportion to its mean value, is larger than that of $D^{SP}$. The differences in credibility regions between FP and SP reflect the ${SP}_{ratio}$, which is 0.78 $\pm$ 0.11 (mean $\pm$ std) during the day, 0.79 $\pm$ 0.12 during the night, and 0.78 $\pm$ 0.10 during the full 24-hour, indicating that the SP is dominant on average.

\subsection{Prediction of Treatment Outcome} \label{Res:corr}
Spearman's rank correlation between the patient-specific $\Delta DV$, as described in Section \ref{sec:pred}, and $\Delta HR$ showed no clear correlation ($p<0.05$) for any combination of drug and AV node property. Hence, no relationship between diurnal variability and drug outcome was found. \newline
\indent ~~~
The Spearman correlation between the patient-specific short-time variability, quantified by $\overline{\Delta KS}$, and $\Delta HR$ showed no clear correlation ($p<0.05$) for the RP and CD in the SP. A moderate correlation was however found between $\overline{\Delta KS}$ and $\Delta HR$ for $R^{FP}$ in the $\beta$-blocker metoprolol ($\rho=0.47, p=0.0011$) and for $D^{FP}$ in metoprolol ($\rho=0.35, p=0.017$). Figure \ref{Figure6} shows the individual $\overline{\Delta KS}$ plotted against $\Delta HR$ and their linear relation for all four drugs, with the left panel showing $R^{FP}$ and the right panel showing $D^{FP}$. Interestingly, a similar relation between $\overline{\Delta KS}$, and $\Delta HR$ is not present in the other $\beta$-blocker carvedilol. \newline
\indent ~~~
The ability to predict $\Delta HR$ using machine learning approaches is evaluated by the average MSE between the predicted and true $\Delta HR$ for the four drugs using the leave-one-out validation method. The average MSE is benchmarked against the population variance of $\Delta HR$ for the four drugs. Hence, if the average MSE is larger than the population variance at 0.0071\%, the population mean yields a more accurate predictor. Using the feature-based regression models, as described in Section \ref{sec:pred}, resulted in an average MSE of 0.0073\% for the linear regression, an average MSE of 0.0074\% for the random forest, and an average MSE of 0.074\% for the k-nearest neighbor. In addition, using the convolutional neural network resulted in an average MSE of 0.0073\% for the fully connected architecture, an average MSE of 0.0079\% for the ResNet architecture, and an average MSE of 0.0074\% for the Inception architecture. Overall, all the machine-learning approaches resulted in an average MSE higher than 0.0071\% and thus in a poor fit to new-seen data.

\section{Discussion} \label{Sec:discussion}
\noindent
A mathematical model with an associated framework for patient-specific estimation and proper uncertainty quantification of the RP and CD in the FP and SP of the AV node using only non-invasive data has been proposed. \newline
\indent ~~~
Individual estimation of trends and variability in AV node properties using non-invasive data has the potential to increase the patient-specific understanding of the AV node during AF, which in turn can be used to enhance informatics approaches for the next generation of personalized medicine. The two most dominant properties of the AV node, the RP and CD, together with the ratio of impulses conducted through the different pathways, have the potential to increase the understanding of the AV node and its function during AF.\newline
\indent ~~~
Due to the physiological differences between the effect of $\beta$-blockers and calcium channel blockers, where $\beta$-blockers reduce the effect of the sympathetic nervous system, our hypothesis was that $\beta$-blockers could exhibit an increased effect when variations in the AV node properties are prominent since this would indicate a larger influence of the autonomic nervous system. The population-averaged trends (Figure \ref{Figure5}) show an increase in RP and a slight decrease in CD during nighttime compared to daytime, suggesting that the decreased sympathetic activity during nighttime affects the RP and CD. However, no correlation was found between diurnal variations in AV properties and reduction in heart rate during treatment with $\beta$-blockers. 
\begin{figure*}[b]
    \centering
        \includegraphics[width=18cm]{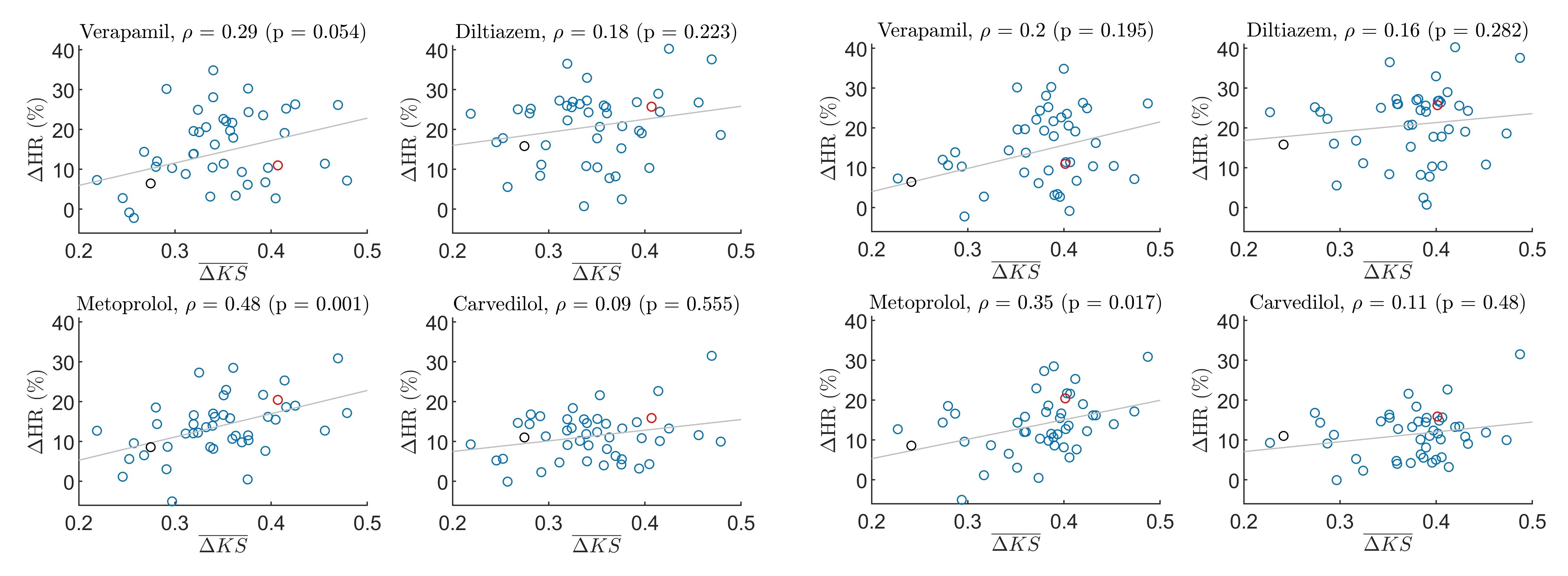}
    \caption{Scatter plot of the 24-hour $\Delta HR$ and $\overline{\Delta KS}$ for the $R^{FP}$ (left) and $D^{FP}$ (right) for the four drugs, with patient A (as shown in Figure \ref{Figure3}) marked with black and patient B (as shown in Figure \ref{Figure4}) marked with red.}
    \label{Figure6}
\end{figure*}
\newline \indent ~~~
However, a potential association between the short-time variability and the treatment outcome with metoprolol was found. The findings depicted in Figure \ref{Figure6} demonstrate a moderate correlation between $\overline{\Delta KS}$ and the change in heart rate ($\Delta HR$) in the RP and CD for the FP for metoprolol, but not for any other drugs. The lack of correlation between $\Delta HR$ after treatment with carvedilol (also a $\beta$-blocker) and $\overline{\Delta KS}$ could potentially be attributed to its modest overall effect observed in the RATAF study, likely stemming from its rapid elimination as acknowledged in \cite{shapiro2013using}. For the possible relationship between short-time variation in the RP and CD in the FP between metoprolol and treatment outcome suggested by this analysis, additional studies are needed to confirm the results. \newline
\indent ~~~
The possibility to predict $\Delta HR$ for the different rate control drugs used in this study was evaluated using three featured-based regression models and three different architectures of a convolutional neural network (Section \ref{Res:corr}). With a resulting average MSE higher than the variance of $\Delta HR$ for the population, it appears impossible to predict $\Delta HR$ with any certainty in the present data set. Either there is not enough information relevant for predicting the heart rate reduction after drug treatment in the AV node property trends -- possibly due to the 10-minute resolution, limiting the information about autonomic regulation -- or the data set size of 51 patients is too low given the inter-individual variability present in the measurements. \newline
\indent ~~~
Prior iterations of the model and framework focused on estimating the model parameter trends rather than the patient-specific property trends of the AV node \cite{karlsson2022ecg}. This approach imposed limitations on the interpretability of the results, since the interpretation of the model parameters in terms of common cardiology terminology such as RP and CD is not straightforward. In contrast, the current work introduces a novel methodology that enables the estimation of the RP and CD for each ECG segment individually, facilitating a more comprehensible and interpretable analysis. The ability to derive such estimates is vital as it allows for effective communication of the analysis results. Furthermore, this advancement in methodology opens up new avenues for gaining a deeper understanding of the AV node and its diurnal and short-term variations. 
\newline
\indent ~~~
The estimation of the posterior by obtaining a range of plausible values, as opposed to relying on a point estimate of the AV node properties, offers notable advantages. For example, the credibility region for $R^{FP}$ in Figure \ref{Figure4} is very broad during most segments at nighttime, reflecting a high uncertainty. In scenarios where the extent of the uncertainty is unknown, these uncertain estimates have the potential to influence decision-making processes or further analysis of the trends. As a result, the usefulness and reliability of these estimates may be decreased, emphasizing the need for a posterior estimation approach. \newline
\indent ~~~
It has previously been shown that a GA can estimate the time-varying network model parameters during 24 hours \cite{karlsson2022ecg}. However, in order to estimate the posterior distribution, the ABC approach was necessary. The ABC approach has in recent years been used for the personalization of the electrophysiological properties in cardiac models \cite{camps2021inference}. Although ABC approaches are generally computationally expensive \cite{turner2012tutorial}, starting in a promising area of the model parameter space, derived from the GA results, reduced the computation time by a factor of around 50 (data not shown). The GA was also used to decide on a reasonable threshold level for the ABC PMC algorithm, which is not straightforward since imperfections in the model make certain RR series more challenging to replicate than others, resulting in a higher average $\epsilon$. Hence, an $\epsilon$ value corresponding to a good fit for one RR interval series could correspond to a poor fit for another, making thresholds very data-dependent. Using the GA to find the threshold levels ensures a reasonable threshold level specified for each data segment.

\subsection{Study Limitations and Future Perspectives}
The estimated RP and CD have not been validated against intracardiac measurements, since obtaining such measurements during AF -- if at all possible -- would be very difficult and time-consuming. The average RP and CD for the two pathways can however be compared with invasive electrophysiological measurements of the AV node from two patients with paroxysmal supraventricular tachycardia and evidence of dual AV nodal conduction found in the literature \cite{denes1973demonstration}. The two patients had an RP in the FP of 820 ms and 495 ms; an RP in the SP of 540 ms and 414 ms; a CD in the FP of 125 ms and 150 ms; and a CD in the SP of 500 ms and 300 ms. Comparing these values to the daytime estimates seen in Table \ref{Table2}, it is evident that the measured values for the RP and CD in both pathways are within the range of our estimated values. It should be noted that the measured functional RP values come from an S1-S2 protocol during sinus rhythm, thus the comparison is not trivial. The functional RP is the smallest AA-interval preceding a conducted impulse. It is however still dependent on the previous pacing frequency, which is not well-defined during AF. Nevertheless, since AF leads to high frequencies, the RP should be reasonably close to the functional RP. In addition, the estimated CD from our model and framework shown in Table \ref{Table2} corresponds to the peak of the probability density function of all CDs in each pathway multiplied by 10. Hence, it differs slightly from the measured CD, since it also captures CDs corresponding to impulses that are blocked within the node. \newline
\indent ~~~
\newline \indent ~~~
In this study, short-time variability was estimated as the difference between adjacent 10-minute intervals. However, limiting the short-time variability to ten minutes also limits the information about the autonomic nervous system -- which is known to operate on a higher resolution -- to a ten-minute resolution. Hence, improving the time resolution of the analysis has the possibility to increase the information extracted by the model and framework, which could improve the results. Furthermore, to extract even more information about the impact of the autonomic nervous system on the AV node, an extension of the model has been proposed in \cite{plappert2022atrioventricular}. A similar framework to the one presented in this work could be employed for that model to estimate model parameters and simulate the RP and CD. This could further refine the estimates and thus the information about the AV node. 
\newline
\indent ~~~
Moreover, analyzing the RP and CD trends for all the patients, a high inter-individual variability with a wide range of diurnal and short-time variability could be seen, likely due to the inherent individual differences. This, in combination with the relatively low number of patients (51), indicates that the results in this paper should be verified in a larger study.

\begin{table*}[b]
\small 
\caption{The mean $\pm$ std of the average $\boldsymbol{\hat{\phi}}_{max}$, the 95\% credibility region, and $\overline{\Delta KS}$ for all patients during daytime, nighttime, and 24-hour average together with $\Delta DV$. The indication $(pat,s)$ is omitted to avoid redundancy. \label{Table2}}
\centering
\begin{tabular}{c c c c c }
\toprule 
& {\scriptsize  $R^{FP}$ }  & {\scriptsize  $R^{SP}$ } & {\scriptsize  $10D^{FP}$ } & {\scriptsize  $10D^{SP}$ }  \\
\cmidrule(rl){2-2}
\cmidrule(rl){3-3}
\cmidrule(rl){4-4}
\cmidrule(rl){5-5}

24-hour $\overline{\boldsymbol{\hat{\phi}}}_{max}$ (ms)
& 934 $\pm$ 203 & 399 $\pm$ 95 & 76.9 $\pm$ 47.6 & 546 $\pm$ 126 \\

Daytime $\overline{\boldsymbol{\hat{\phi}}}_{max}$ (ms)
& 839 $\pm$ 205 & 356 $\pm$ 94 & 85 $\pm$ 64.6 & 572 $\pm$ 139 \\

Nighttime $\overline{\boldsymbol{\hat{\phi}}}_{max}$ (ms)
& 1119 $\pm$ 294 & 481 $\pm$ 152 & 62.1 $\pm$ 52.8 & 484 $\pm$ 160 \\

24-hour $\overline{ \boldsymbol{\hat{\phi}}}_{95} - \overline{\boldsymbol{\hat{\phi}}}_{5}$ (ms)
& 687 $\pm$ 232 & 217 $\pm$ 114 & 304.1 $\pm$ 110.7 & 447 $\pm$ 103 \\

Daytime $\overline{ \boldsymbol{\hat{\phi}}}_{95} - \overline{\boldsymbol{\hat{\phi}}}_{5}$ (ms)
& 671 $\pm$ 261 & 179 $\pm$ 103 & 299.4 $\pm$ 123.9 & 427 $\pm$ 94 \\

Nighttime $\overline{ \boldsymbol{\hat{\phi}}}_{95} - \overline{\boldsymbol{\hat{\phi}}}_{5}$ (ms)
& 738 $\pm$ 290 & 291 $\pm$ 185 & 315.5 $\pm$ 153.3 & 477 $\pm$ 169 \\

24-hour $\overline{\Delta KS}$
& 0.347 $\pm$ 0.057 & 0.319 $\pm$ 0.136 & 0.376 $\pm$ 0.055 & 0.36 $\pm$ 0.07 \\

Daytime $\overline{\Delta KS}$
& 0.368 $\pm$ 0.069 & 0.352 $\pm$ 0.169 & 0.393 $\pm$ 0.061 & 0.351 $\pm$ 0.089 \\

Nighttime $\overline{\Delta KS}$
& 0.309 $\pm$ 0.083 & 0.253 $\pm$ 0.133 & 0.342 $\pm$ 0.075 & 0.38 $\pm$ 0.082 \\

$\Delta DV$ 
& 0.77 $\pm$ 0.18 & 0.78 $\pm$ 0.27 & 2.58 $\pm$ 3.72 & 1.29 $\pm$ 0.47 

\\\bottomrule
\end{tabular}
\end{table*}

\section{Conclusion} \label{Sec:conclusion}
 \label{Sec:conclusion}
\noindent 
We have proposed a novel framework for estimating patient-specific 24-hour trends of the RP and CD in the FP and SP of the AV node by mapping estimated model parameters. These estimates include the full posterior of the RP and CD and could be estimated using only non-invasive data. 
Additionally, a correlation between short-term variability in both the RP and CD for the FP and drug-induced changes to the heart rate was found. The individual estimates of AV node properties offer patient-specific trends in RP and CD, which may have the potential to assist in treatment selection.

\newpage
\section{Conflict of Interest Statement}
The authors declare that the research was conducted in the absence of any commercial or financial relationships that could be construed as a potential conflict of interest.

\section{Author Contributions}
\noindent MK, FS, and MW contributed to the design and conception of the study. SU performed the clinical study. FS was responsible for estimating the RR interval series and AFR trends from the ECG. MK wrote the manuscript, designed the genetic algorithm, the approximate Bayesian computation algorithm, and the model reduction, with advice, suggestions, and supervision from FS and MW. SU and PP analyzed and interpreted the results from a medical viewpoint. FS and MW supervised the project and reviewed the manuscript during the writing process. All authors contributed to the manuscript revision, read, and approved the submitted version.

\section{Funding}
\noindent This work was supported by the Swedish Foundation for Strategic Research (Grant FID18-0023), the Swedish Research Council (Grant VR2019-04272), and the Crafoord Foundation (Grant 20200605).



\section{Data Availability Statement} \label{Sec:DAS}
\noindent The estimated AV node properties $\boldsymbol{\hat{\Phi}}(pat,s)$ supporting the conclusions for this article will be available from MK upon request. The measured data are owned by Vestre Viken Hospital Trust, and requests for access can be made to SU. The code for the model together with a user example can be found at \url{https://github.com/FraunhoferChalmersCentre/AV-node-model}.

\bibliographystyle{IEEEtranN}
\bibliography{Ref}

\end{document}